\documentclass[fleqn,usenatbib]{mnras}
\usepackage{newtxtext,newtxmath}
\usepackage[T1]{fontenc}
\usepackage{graphicx}
\usepackage{amsmath}
\usepackage{threeparttable, tablefootnote}
\usepackage{booktabs}

\newcommand{\angstrom}{\mbox{\normalfont\AA}}
\newcommand{\kms}{\hbox{km\,s$^{-1}$}}

\title[Blazar Redshift Constraints]{Improving blazar redshift constraints with the edge of the Ly$\alpha$ forest: 1ES 1553+113 and implications for observations of the WHIM}

\author[J. Dorigo Jones et al.]{
J. Dorigo Jones,$^{1,2}$\thanks{E-mail: johnny.dorigojones@colorado.edu}
S. D. Johnson,$^{1}$\thanks{E-mail: seanjoh@umich.edu}
Sowgat Muzahid,$^{3,4}$
J. Charlton,$^{5}$
H.-W. Chen,$^{6}$
A. Narayanan,$^{7}$
\newauthor Sameer,$^{5}$
J. Schaye,$^{8}$
N. A. Wijers$^{8}$
\\
$^{1}$Department of Astronomy, University of Michigan, 1085 South University Ave., Ann Arbor, MI 48109-1107, USA\\
$^{2}$Department of Astrophysical and Planetary Sciences, University of Colorado Boulder, 391 UCB, 2000 Colorado Ave., Boulder, CO 80309, USA\\
$^{3}$IUCAA, Post Bag-04, Ganeshkhind, Pune, India - 411007\\
$^{4}$Leibniz-Institut fur Astrophysik Potsdam (AIP), An der Sternwarte 16, 14482 Potsdam, Germany\\
$^{5}$Department of Astronomy \& Astrophysics, The Pennsylvania State University, 525 Davey Lab, University Park, PA 16802, USA\\
$^{6}$Department of Astronomy \& Astrophysics, The University of Chicago, 5640 South Ellis Avenue, Chicago, IL 60637, USA\\
$^{7}$Indian Institute of Space Science and Technology, Thiruvananthapuram 695547, Kerala, India\\
$^{8}$Leiden Observatory, Leiden University, PO Box 9513, NL-2300 RA Leiden, The Netherlands
}

\date{Accepted 2021 November 10. Received 2021 November 1; in original form 2021 August 27}

\pubyear{2021}

\begin{document}
\label{firstpage}
\pagerange{\pageref{firstpage}--\pageref{lastpage}}
\maketitle

\begin{abstract} 
Blazars are some of the brightest UV and X-ray sources in the sky and are valuable probes of the elusive warm-hot intergalactic medium (WHIM; $T{\simeq} 10^5{\textrm{--}}10^7$ K). However, many of the brightest blazars--called BL Lac objects such as 1ES1553+113--have quasi-featureless spectra and poorly constrained redshifts. Here, we significantly improve the precision of indirect redshift constraints for blazars based on the edge of the \ion{H}{i} Ly$\alpha$ forest observed in their UV spectra. We develop a robust technique to constrain the redshift of a $z<0.5$ AGN or blazar with a $1\sigma$ uncertainty of ${\approx}0.01$ using only the position of its highest-redshift Ly$\alpha$ absorber with $\log N_{\rm{\ion{H}{i}}}/{\rm cm^{-2}} > 12.6$. We use a large sample of 192 AGN/QSOs at $0.01\lesssim z\lesssim0.45$ that have high-quality COS FUV spectra to characterize the intrinsic scatter in the gap between the AGN redshift and the edge of their Ly$\alpha$ forest. We present new COS NUV data for 1ES1553+113 and confirm its redshift of $z=0.433$ using our technique. We apply our Ly$\alpha$-forest-based redshift estimation technique to nine additional blazars with archival \textit{HST} UV spectra, most of which are key targets for future X-ray missions. Our inferred redshift constraints improve estimates for two BL Lacs (1ES1118+424 and S50716+714) and are consistent with previous estimates for the rest. Our results emphasize the need to obtain further UV spectra of bright blazars, of which many have uncertain redshifts, in order to maximize the scientific value of future X-ray WHIM observations that will improve our understanding of galaxy evolution.
\end{abstract}

\begin{keywords}
(galaxies:) intergalactic medium -- (galaxies:) quasars: absorption lines -- (galaxies:) BL Lacertae objects: individual: 1ES 1553+113 -- (galaxies:) BL Lacertae objects: general
\end{keywords}

\section{Introduction}
\label{intro}
A number of active galactic nuclei (AGN) have a relativistic jet that is oriented directly toward Earth, and the resulting beaming causes them to be some of the brightest objects observed in the Universe. These extreme AGN, known as blazars, have been the catalyst for many discoveries in high-energy astrophysics and observational cosmology. Blazars can be extraordinarily bright in the ultraviolet (UV), X-ray, and even $\gamma$-ray, making them ideal background sources for absorption studies of the diffuse intergalactic medium (IGM; see reviews by \citealt[][]{Bregman07, McQuinn16}) and also for constraints on the extragalactic background light (EBL) (see, e.g., \citealt[][]{Stecker07, BiteauWilliams} and the review by \citealt[][]{MadejskiSikora}).

Blazars are particularly promising probes of the elusive warm-hot phase of the IGM (WHIM; $T \simeq 10^5 \textrm{--} 10^7$ K), which cosmological simulations predict contains up to $\approx$50 per cent of all baryons at low-redshift ($z \lesssim 1$; \citealt[][]{CenOstriker99, CenOstriker06, Cui19, Martizzi19, Tuominen21}), with the other half of baryons residing in galaxies \citep[][]{FukugitaPeebles}, the circumgalactic medium (CGM; see review by \citealt[][]{Tumlinson17}), the intracluster or intragroup medium (ICM or IGrM; \citealt[][]{Simionescu11, Urban11, Eckert15}), and the cooler photoionized IGM ($T \simeq 10^4 \textrm{--} 10^5$ K; \citealt[][]{Penton04, Shull12}). 

Detecting the WHIM beyond the local group \citep[e.g.,][]{Bregman07b, Anderson10, Gupta12, Miller13, Fang15, Qu21} has been a significant observational challenge because of its low average density ($n_{\rm H} \lesssim 10^{-4}$ cm$^{-3}$) and high temperatures, leading to what is known as the `missing baryons' problem at low-redshift. Recently, the localisation of fast radio bursts (FRBs) to their host galaxies has enabled the likely closure of the missing baryon problem through the excess dispersion measure attributed to free electrons in the IGM \citep[][]{Ravi19, Bannister19, Macquart20, Ocker21}; however, there are only a few such localized FRBs \citep[][]{Walker20}, and the dispersion contribution from the WHIM relative to the cool IGM is difficult to constrain directly (e.g., \citealt[][]{Shull18, MedlockCen21}) even when local and FRB host contributions are controlled for. This leaves important questions regarding the prevalence, distribution, phase, and metallicity of the WHIM largely unanswered. A more complete characterization of the WHIM is necessary not only as a check of our understanding of cosmic structure formation, but also to gain insight into galactic feedback and the flow of baryons through the CGM and IGM (e.g., \citealt[][]{SomervilleDave15, AnglesAlcazar17, Peeples19, Wijers19, Wijers20, MitchellSchaye21}).

So far, only the lower temperature regime ($T \simeq 10^5\textrm{--}10^{5.5}$ K) of the WHIM has been observed in large spectroscopic samples: through broad neutral hydrogen Lyman~$\alpha$ (\ion{H}{i} Ly$\alpha$) and \ion{O}{vi} absorption in the far-UV (FUV) \citep{Tripp08, ThomChen08, Danforth10, Narayanan10, Tilton12, Shull12, Danforth16}. In smaller samples, \ion{Ne}{viii} absorption has been used to detect warm-hot baryon reservoirs primarily at $z \gtrsim 0.5$ because of the available UV coverage \citep[][]{Meiring13, Pachat17, Frank18, Burchett19}. At somewhat higher temperatures ($T \gtrsim$ few $\times 10^6$ K), the WHIM has been probed using the thermal Sunyaev--Zel’dovich (tSZ) effect around massive galaxies and clusters \citep{Tanimura19, deGraaff19, Singari20}. These detections can account for about two-thirds of the expected baryon content of the WHIM, leaving $\approx$18 per cent of all baryons at low-redshift still elusive \citep[][]{deGraaff19}.

As such, the exhaustive search for and characterization of the hot phase of the WHIM currently relies on difficult-to-detect atomic transitions from highly-ionized metals expected toward X-ray bright sightlines that probe $T \gtrsim 10^6$ K gas. Some of the strongest expected tracers of the hot WHIM are intervening \ion{O}{vii} $\lambda 21.6\ \angstrom$ and \ion{O}{viii} $\lambda 18.9\ \angstrom$ absorption lines in the X-ray \citep[e.g.,][]{Hellsten98, Wijers19}. Possibilities exist in the FUV, such as the \ion{O}{vi} $\lambda \lambda1031, 1037\ \angstrom$ and \ion{Ne}{viii} $\lambda \lambda770, 780\ \angstrom$ doublets \citep[e.g.,][]{Mulchaey96}, but these ions are not unambiguous tracers of hot gas as they can be either photoionized or collisionally-ionized \citep[e.g.][]{Tripp08, ThomChen08, Savage14, Hussain17, Narayanan18}, and they would be harder to detect than \ion{O}{vii} or \ion{O}{viii} at these temperatures given their decreasing ion fractions over the temperature range $T \simeq 10^{5.5}\textrm{--}10^{6}$ K \citep[e.g.,][]{OppenheimerSchaye13}.

Detections of the hot WHIM toward blazars in X-ray absorption have so far been contested, mainly as a result of the limited sensitivity and spectral resolution of available X-ray telescopes (\citealt[][]{Fang02, Nicastro05a, Nicastro05b, Kaastra06, Bregman07, Fang10, Ren14, Nicastro16, Bregman18}). Recent X-ray follow-up studies of the UV-detected WHIM utilized stacking to obtain tentative detections of intervening \ion{O}{vii} or \ion{O}{viii} absorption \citep[][]{Bonamente16, Nevalainen19, Kovacs19, Ahoranta20}, although the physical properties (e.g., metallicity, ionization state) of the absorbing gas are uncertain \citep[see][and references therein]{Bregman19}. Indeed, direct detections of the hot WHIM in individual absorbers are currently limited to a few systems, but this will be revolutionized by future X-ray missions \citep[see][]{Wijers20}, such as \textit{Athena} (approved, planned launch early 2030s, \citealt{Athena}), \textit{Arcus} (proposed, \citealt[][]{Arcus}), and \textit{Lynx} (concept, \citealt[][]{Lynx}). This underscores the need to prepare sufficient sightlines of distant blazars that have well-determined redshifts and that are bright enough to probe the IGM via both UV and X-ray absorption spectroscopy \citep[e.g.,][]{Bregman15}.

To probe the hot WHIM in absorption over the largest available redshift pathlength, \citet[][]{Nicastro18a} observed the bright, featureless blazar 1ES 1553+113 using 1.75 Ms of \textit{XMM-Newton} grating spectroscopy. The exceptionally high signal-to-noise (S/N) X-ray spectrum of 1ES 1553+113 revealed two candidate \ion{O}{vii} absorption systems along the line-of-sight at $z =$ 0.4339 and 0.3551 that were thought to arise from the WHIM. A census based on these candidate detections can account for the remaining `missing baryons' \citep[][]{Nicastro18b}.

However, 1ES 1553+113 has no direct spectroscopic redshift measurement because its spectrum lacks any detected intrinsic absorption or emission lines. In order to place a limit on the redshift of 1ES 1553+113, \citet[][]{Danforth10} utilized the presence of intervening absorption in its FUV spectrum. They inferred a redshift constraint for 1ES 1553+113 of $0.433 < z < 0.58$ by comparing the observed and expected \ion{H}{i} Ly$\alpha$ line densities. This was later corrected by \citet[][]{Danforth16} to be $0.413 < z < 0.56$ after accounting for a previous line misidentification. Unfortunately, the large range in possible redshifts rendered the origin of the $z =$ 0.4339 candidate \ion{O}{vii} absorber ambiguous.

In efforts to pin down the systemic redshift of 1ES 1553+113 and study absorber origins, \citet[][]{PaperI} performed a deep spectroscopic redshift survey toward this blazar and inferred a statistical redshift constraint from its \ion{H}{i} Ly$\alpha$ forest and group environment. They found that 1ES 1553+113 is most likely a member of a galaxy group identified at $z =$ 0.433. However, confirmation of this result requires further spectral coverage toward 1ES 1553+113 in the near-UV (NUV) and a more robust characterization of the edge of the observed Ly$\alpha$ forest in order to unambiguously determine the redshift of this blazar.

In fact, many bright blazars, such as 1ES 1553+113, have no measured spectroscopic redshift. These quasi-featureless blazars, known as BL Lac objects, comprise approximately half of all known blazars, with the rest known as flat-spectrum radio quasars (FSRQs) \citep[e.g.,][]{Healey07, Falomo14, Mao17}. Much work has been done to constrain the redshifts of BL Lac objects through both direct and indirect methods, such as detecting intrinsic but weak absorption or emission features from the blazar's host galaxy or nucleus \citep[e.g.,][]{Stocke11, Fang14, vandenBosch15, Danforth16, Archambault16, Paiano17, Furniss19, Bu19, Goldoni21, Paiano21}, associating the blazar with a nearby galaxy group \citep[e.g.,][]{Farina16, Rovero16, Torres-Zafra18, PaperI}, or inferring a redshift constraint from the blazar's highest-redshift \ion{H}{i} Ly$\alpha$ absorption line seen in the UV (e.g., \citealt[][]{Blades85, Danforth10, Danforth13, Furniss13a, Furniss13b, PaperI}). For blazars at high-redshift ($z \gtrsim 1.3$), there is an additional redshift constraint tool in the photometric dropout method \citep[][]{Rau12, Kaur17, Rajagopal20}.

Accurate redshift constraints for featureless blazars are essential for interpreting studies of the WHIM in order to classify detected absorption systems as arising from either the IGM or the blazar's group environment, the latter of which cannot be used for cosmological measurements. Studies of the EBL conducted toward bright blazars like 1ES 1553+113 also rely on well-constrained redshifts. In addition, Ly$\alpha$ forest based redshift constraints may be useful for studying other aspects of blazars, such as the unique class of gravitationally lensed blazars (e.g., \citealt[][]{OstrikerVietri85, Stickel88, Falomo92, Readhead21}) in order to avoid confusion with features from the lensing galaxy.

Our goal in this paper is twofold:
\begin{enumerate}
\item verify the systemic redshift of 1ES 1553+113 with new NUV spectroscopy to better understand the origin of the X-ray absorbers detected toward this blazar, and 
\item improve the quantification and precision of indirect redshift measurements for blazars based on the edge of the \ion{H}{i} Ly$\alpha$ forest in their spectra.
\end{enumerate} 

This paper proceeds as follows: In Section~\ref{sec:observations}, we present new NUV observations of the blazar 1ES 1553+113 and confirm the extent of its Ly$\alpha$ forest. In Section~\ref{sec:redshift}, we constrain the redshift of 1ES 1553+113 using its highest-redshift Ly$\alpha$ line in combination with a more complete characterization of the edge of the \ion{H}{i} Ly$\alpha$ forest observed in archival UV spectra of AGN at $z \lesssim 0.45$. We use this archival sample to evaluate statistical and systematic uncertainty in the redshift technique and compare to previous methods. In Section~\ref{sec:application}, we apply our redshift constraint technique to other well-known blazars and demonstrate its utility for future studies of the WHIM and for constraints on the EBL. We briefly summarize our conclusions in Section~\ref{sec:conclusions}.

\section{NUV Observations and Spectral Analysis of 1ES 1553+113}
\label{sec:observations}

\begin{figure*}
    \includegraphics[width=\linewidth]{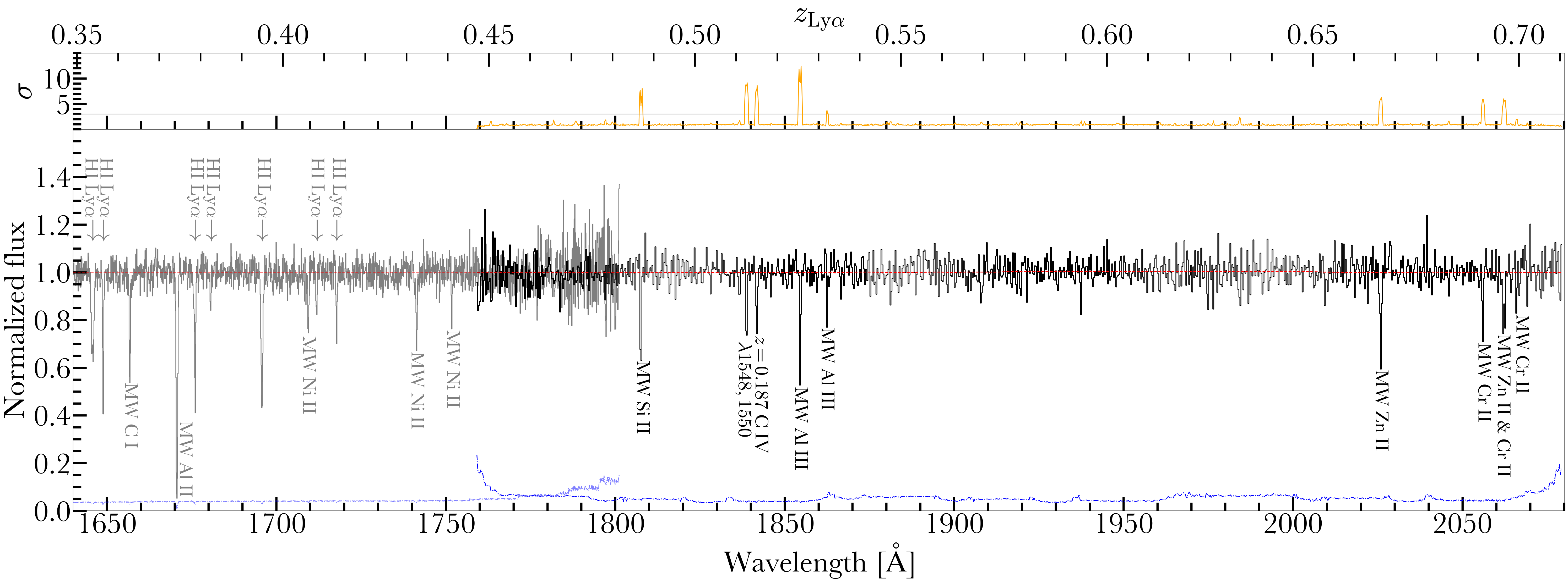}
    \caption{Co-added \textit{HST}/COS NUV G185M spectrum of the blazar 1ES 1553+113 binned by 8 pixels, with normalized flux shown in black, error in blue, and the continuum in red. The bottom axis shows the observed-frame wavelength and the top axis shows the corresponding Ly$\alpha$ redshift. Intervening \ion{H}{i} Ly$\alpha$ absorption systems and MW features are labeled, and the estimated significance of absorption line detection throughout the NUV spectrum is shown in the top panel in orange with the 3$\sigma$ significance level indicated by the horizontal gray line (see Section~\ref{sec:observations}). The redward part of the COS FUV G160M spectrum of 1ES 1553+113 is shown in gray \citep[][]{Danforth10, PaperI}, which contains the highest-redshift \ion{H}{i} Ly$\alpha$ line seen toward this blazar (see Section~\ref{sec:observations}).}
    \label{fig:spectrum}
\end{figure*}

As summarized in the introduction, the redshift of 1ES 1553+113 must be constrained through indirect methods, because its intrinsic spectral features are currently undetectable. The highest-redshift \ion{H}{i} Ly$\alpha$ absorption line seen toward 1ES 1553+113 in the FUV at $z = 0.413$ has been used as a redshift lower limit to infer the blazar's membership with a galaxy group at $z = 0.433$, thus preventing the use of the $z = 0.4339$ \ion{O}{vii} candidate absorber in censuses of the WHIM \citep[][]{PaperI}. However, given the position of the highest-redshift Ly$\alpha$ line detected relative to the end of the FUV spectrum (see Fig.~\ref{fig:spectrum}), it is possible that a higher-redshift Ly$\alpha$ line could exist at $z \gtrsim 0.46$, seen in the NUV, which would invalidate the inferred redshift constraint.

Because of this, following \citet[][]{PaperI}, we obtained high-quality NUV spectra of 1ES 1553+113 with the Cosmic Origins Spectrograph (COS; \citealt{Green12}) on the \textit{Hubble Space Telescope} (\textit{HST}) to search for \ion{H}{i} Ly$\alpha$ absorption at 0.45 $\lesssim z \lesssim$ 0.71 (PI: Muzahid, PID: 15835, \citealt[][]{Muzahid19}). The absence of any such Ly$\alpha$ absorption in the NUV would help to verify the galaxy group association inferred for 1ES 1553+113, while detection of additional Ly$\alpha$ absorption in the NUV would imply that the blazar is located at higher redshift.

We acquired the NUV spectroscopy of 1ES 1553+113 over four orbits with the COS G185M grating using four central wavelength (CENWAVE) settings: 1882, 1913, 1941, and 1953 $\angstrom$, each observed for one orbit, for a total of 8.7 ks of COS NUV spectroscopy. We split the observations in each CENWAVE setting into four exposures with different grating offset positions (FP-POS $=$ 1--4) to minimize flat fielding errors. The resulting 16 exposures provide contiguous spectral coverage from $1759\,\angstrom$ to $2079\,\angstrom$, including some overlap with the COS FUV spectrum of 1ES 1553+113.

We performed the initial data reduction of these NUV observations using default parameters in the COS calibration pipeline (\textsc{CalCOS}, v3.3.9) to generate one-dimensional extracted spectra. The \textsc{x1dcorr} module in \textsc{CalCOS} performs a BOXCAR spectral extraction on each NUV spectral stripe (i.e. NUVA, NUVB, NUVC) in the two-dimensional spectral image (i.e. flat-fielded image file) using values for the slope, y-intercept, and extraction box height (EBH) of each stripe for each CENWAVE setting (see \citealt[][]{Snyder})\footnotemark
\footnotetext{Also see the recently released `COS Walkthrough Notebooks' by Space Telescope Science Institute (STScI): https://github.com/spacetelescope/notebooks/tree/master/notebooks/COS}. However, the default extraction values do not maximize the S/N in the one-dimensional spectra, primarily because the default EBH of 57 pixels was chosen to minimize flux loss instead.

To improve the S/N in the NUV spectrum of 1ES 1553+113, we modified the spectral extraction aperture used in the data reduction process. We fit the trace along each spectral exposure to update the y-intercept and slope values, and we ran a grid of EBH values (7 $\leq$ EBH $\leq$ 23) to determine the optimal EBH. We decided to use EBH $=$ 15 pixels for each stripe because this value maximizes the S/N by decreasing the noise from dark current without significant loss of signal. Finally, we removed pixels with data quality (DQ) flag $>$ 4. These modifications improved the S/N per resolution element (resel) in the reduced spectra by 10--15 per cent compared to the default spectral extraction, resulting in a median S/N per resel of $\approx$12.

We then continuum-normalized and median-combined the resulting 16 modified one-dimensional spectra to produce the final co-added NUV spectrum of 1ES 1553+113, displayed in Fig.~\ref{fig:spectrum}. We note that the first orbit for the 1953 CENWAVE setting failed because of a guide star acquisition issue, and we subsequently obtained repeat observations at this CENWAVE setting three months later. Between these observations, 1ES 1553+113 decreased in NUV flux by $\approx$25 per cent, which is unsurprising given its known variability \citep[e.g.,][]{Pandey19, Dhiman21}. This dimming required us to continuum-normalize the individual stripe co-adds for each CENWAVE setting, using a low-order polynomial fit, before median-combining them to create the final co-added spectrum. We also confirmed that the \textsc{CalCOS} wavelength calibration (wavecal) is consistent with the expected accuracy of 15 km s$^{-1}$ for the NUV channel by measuring the centroids of low ionization Milky Way (MW) transitions.

We observed no detectable \ion{H}{i} Ly$\alpha$ emission line in the archival FUV and new NUV spectra of 1ES 1553+113, and so we proceeded to search for \ion{H}{i} Ly$\alpha$ absorption that could be used as an indirect redshift constraint. We visually identified the absorption features present in the high-quality NUV spectrum of 1ES 1553+113 over the wavelength range $1759\,\angstrom\,<\,\lambda\,<\,2079\,\angstrom$, corresponding to $0.45 \lesssim z_{\rm Ly\alpha} \lesssim 0.71$. The spectrum contains absorption at $z \approx$ 0 from four low-ionization MW metal ions (\ion{Si}{ii}, \ion{Al}{iii}, \ion{Zn}{ii}, and \ion{Cr}{ii}), and also a \ion{C}{iv} $\lambda \lambda 1548, 1550 \angstrom$ doublet arising from an absorbing system at $z \approx 0.187$ (see Fig.~\ref{fig:spectrum}). This intervening absorber is also detected in \ion{O}{vi}, \ion{N}{v}, \ion{C}{iii}, and \ion{H}{i} in the FUV \citep[][]{Danforth10}, and ionization analysis of this system will be presented in a future work.

To verify these detections, characterize the sensitivity of the data, and identify any possible additional absorption features, we performed a blind search for absorption lines by quantifying our ability to detect absorption throughout our NUV spectrum. Our line-finding procedure steps through the spectrum on a fine grid in wavelength ($\Delta z = 0.0001$, corresponding to typical \ion{H}{i} Ly$\alpha$ absorber line widths of $\approx$20 km s$^{-1}$ at $z\approx 0.4$ \citep[e.g.,][]{Danforth16}) and calculates the best-fitting Voigt profile and the corresponding line strength, line width, and associated uncertainties using a Markov Chain Monte--Carlo (MCMC) method. Specifically, at each step we fit a Voigt profile with initial guesses for the line strength (i.e. column density multiplied by the transition cross-section: $N \times \frac{\pi e^2}{m_e c} f_{jk}$, where $f_{jk}$ is the oscillator strength) and Doppler width ($b$) that correspond to a typical \ion{H}{i} Ly$\alpha$ line (initial guesses: $\log N_{\rm{\ion{H}{i}}}/{\rm cm^{-2}} = 13.5$ and $b = 15$ km s$^{-1}$). The best-fitting values for line strength, $b$, and the centroid are calculated at each step by maximizing the likelihood function. The likelihood function is combined with our priors (10 $\kms < b <$ 50 $\kms$, $\log N_{\rm{\ion{H}{i}}}/{\rm cm^{-2}} >$ 11) to create the full $\log$ likelihood function, which is then sampled at each wavelength step via MCMC using the \textsc{emcee} module \citep[][]{emcee} for a total of 3000 steps with a burn-in period of 2000 steps. We then computed the statistical significance, $\sigma$, of any absorption line detection as a function of wavelength in our NUV spectrum, shown in the top panel in Fig.~\ref{fig:spectrum}.

This blind search recovers our manual line identifications at $\sigma \gtrsim 3$ and does not return any additional lines. The 3$\sigma$ detection limit corresponds to a minimum detected \ion{H}{i} column density in our NUV spectrum of $\log N_{\rm{\ion{H}{i}}}/{\rm cm^{-2}} \gtrsim 12.6$. The only manual line identification not recovered at $\sigma \gtrsim 3$ is the weaker MW \ion{Cr}{ii} line, but we note that this is indeed a commonly observed line in high \ion{H}{i} column density systems \citep[e.g.,][]{Mas-Ribas17}. 

Most notably, we find no new \ion{H}{i} Ly$\alpha$ absorption in the NUV. This confirms the lower limit constraint on the blazar's redshift of $z_{\rm{sys}} \gtrsim$ 0.413 based on the position of its highest-redshift \ion{H}{i} Ly$\alpha$ line in the FUV, seen in Fig.~\ref{fig:spectrum} \citep[][]{Danforth16, PaperI}. In order to infer a redshift upper limit constraint for 1ES 1553+113, or any blazar, based on its highest-redshift Ly$\alpha$ line, in the following section we closely examine the edge of the Ly$\alpha$ forest observed toward low-redshift AGN with systemic redshifts available in the literature.

\section{Redshift constraint based on the edge of the observed \texorpdfstring{\ion{H}{i}}{H I} \texorpdfstring{Ly$\alpha$ forest}{Lya}}
\label{sec:redshift}

In this section, we develop and explore systematics of AGN and blazar redshift constraints based on the highest-redshift \ion{H}{i} Ly$\alpha$ absorption line observed in their spectra. In Section~\ref{subsec:1ESredshift}, we apply this redshift constraint technique to 1ES 1553+113 and confirm its inferred galaxy group association. In Sections~\ref{subsec:systematics} and~\ref{subsec:evolution} we discuss the systematics of our technique, and in Section~\ref{subsec:comparison} we compare it to previous methods.

\subsection{AGN redshift constraints from the \texorpdfstring{Ly$\alpha$}{Lya} forest at \texorpdfstring{$z \lesssim 0.45$}{z<0.45}}
\label{subsec:forest}

To develop our redshift constraint technique, we carry out a robust characterization of the edge of the \ion{H}{i} Ly$\alpha$ forest in a large sample of 192 AGN at $0.01\,\lesssim\,z\,\lesssim\,0.45$ with FUV spectra available in the COS archive. The redshift upper limit of $z\,\lesssim\,0.45$ is chosen to ensure COS coverage of the entire Ly$\alpha$ forest toward the AGN. Each of these AGN has a spectroscopic redshift measured from intrinsic emission lines, which enables us to construct an empirical cumulative distribution function of the difference between the AGN systemic redshifts and that of the highest-redshift Ly$\alpha$ absorption line in their UV spectra.

We construct our sample of Ly$\alpha$ forest sightlines by retrieving the public data for all of the 584 AGN/quasi-stellar objects (QSOs) in the Hubble Spectroscopic Legacy Archive (HSLA; \citealt{HSLA}) that have COS FUV coverage in the G130M or G160M gratings (as of October, 2020). We discarded all objects that have median S/N per resel $<$ 10 within 5 $\angstrom$ of the Ly$\alpha$ emission line to ensure a detection limit comparable to or better than our NUV spectrum of 1ES 1553+113 (see Section~\ref{sec:observations} and below), leaving 205 AGN. We dropped nine of these AGN because they were targeted to study broad associated absorption lines (BALs or mini-BALs) from outflows, which could skew the sample away from typical AGN. One object was dropped due to significant contamination from the MW Ly$\alpha$ damping wing. We also dropped three objects that are blazars because we later apply the resulting redshift constraint to these objects (see Section~\ref{subsec:consistent}). The final sample consists of 192 AGN/QSOs (see Table~\ref{tab:data}), located at $z \lesssim 0.45$, which we use to characterize the edge of the observed \ion{H}{i} Ly$\alpha$ forest relative to the systemic redshift of the AGN.

We retrieved systemic redshifts, $z_{\rm{sys}}$, for the sample from the literature by first finding matches in the Sloan Digital Sky Survey (SDSS) Quasar Catalogue \citep[][]{SDSSDR5Q, SDSSDR12Q, SDSSDR16Q}, then in the UV-Bright Quasar Survey (UVQS) DR1 \citep[][]{UVQS}, then in the 13$^{\rm th}$ edition of the Catalogue of Quasars \& AGN by V{\'e}ron-Cetty and V{\'e}ron (VCV; \citealt[][]{Veron}). For the 74 objects with no match in SDSS or UVQS, we used the most accurate spectroscopic redshift given in the NASA/IPAC Extragalactic Database (NED) if more accurate than the redshift from VCV. In the final sample, 95 objects use $z_{\rm{sys}}$ from SDSS, 27 from VCV, 23 from UVQS, 13 from the 6dF Galaxy Survey DR3 \citep[][]{6dF}, and 34 from various other sources\footnotemark
\footnotetext{\citet[][]{Allen78, Bergeron83, Schmidt83, Hintzen84, Remillard86, Dressler88, deVaucouleurs91, Fouque92, Huchra93, Corwin94, Marziani96, Pietsch98, Huchra99, Falco99, Wisotzki00, Springob05, Ho09, Huchra12}}. While the quality of the literature redshifts varies, they are all based on rest-frame optical emission lines, such as the Balmer series, [\ion{O}{ii}], [\ion{O}{iii}], and \ion{Mg}{ii}, which have typical systematic uncertainties of $\approx$75 km s$^{-1}$, based on a comparison between \ion{Mg}{ii} and [\ion{O}{iii}] by \citet[][]{HewettWild10}.

We identified the highest-redshift Ly$\alpha$ line, max($z_{\rm{Ly\alpha}}$), in each AGN spectrum by fitting a single-component Voigt profile, making sure not to choose $z \approx$ 0 MW lines. In order to avoid misidentifying a $z>0$ metal line as \ion{H}{i} Ly$\alpha$, we tested other possible line identifications and searched for the corresponding \ion{H}{i} Ly$\alpha$ line at the same redshift. This process confirms confident max($z_{\rm{Ly\alpha}}$) identifications, except in a few ambiguous cases, which do not affect our results (HB89-0202-765, SDSS-J104741.75+151332.2, and SDSSJ213357.89-071217.3, see Table~\ref{tab:data}). For AGN with max($z_{\rm{Ly\alpha}}$) at $z\,\gtrsim\,0.15$, we confirmed the detection with the corresponding Ly$\beta$ line. Also, for 54 AGN ($\approx$28 per cent of the sample), we confirmed our max($z_{\rm{Ly\alpha}}$) by comparing with line identifications from \citet[][]{Danforth16} \footnotemark
\footnotetext{https://doi.org/10.17909/T95P4K}. We required the max($z_{\rm{Ly\alpha}}$) to have $\log N_{\rm{\ion{H}{i}}}/{\rm cm^{-2}} >$ 12.6 and spectral S/N per resel $>$ 10 in order to obtain $3\sigma$ detections\footnotemark.
\footnotetext{Even though our line selection criteria for max($z_{\rm{Ly\alpha}}$) reflect the 3$\sigma$ confidence limit for detecting absorption lines in our NUV spectrum of 1ES 1553+113, it is important to ensure that the chosen S/N cut and minimum $\log N_{\rm{\ion{H}{i}}}$ are consistent such that we can confidently detect and measure unblended Ly$\alpha$ stronger than the chosen limit in the AGN spectra. To test this, we calculated the median 3$\sigma$ limit for absorption line detection (see Section~\ref{sec:observations} for description of procedure) in the regions between max($z_{\rm{Ly\alpha}}$) and $z_{\rm{sys}}$ for the 16 AGN with the lowest S/N spectra (i.e. S/N per resel $<$ 20--30 near the Ly$\alpha$ emission) and also for the 10 AGN with max($z_{\rm{Ly\alpha}}$) most separated from the AGN (i.e. $\Delta v = [c z_{\rm{sys}} - c$ max($z_{\rm{Ly\alpha}})]/(1+z_{\rm{sys}}) >$ 3000 $\kms$). We found that only five of the low-S/N objects have a 3$\sigma$ detection limit slightly worse than our NUV spectrum of 1ES 1553+113 (i.e. minimum detected $\log N_{\rm{\ion{H}{i}}}/{\rm cm^{-2}} \gtrsim 12.6$). All of the other AGN tested have a 3$\sigma$ detection limit better than our NUV spectrum, and the remaining, higher-S/N objects in our sample that were not tested would have even better detection limits. As such, these five AGN are included as right-censored data points, as described in the text.}

\begin{figure*}
    \includegraphics[width=\linewidth]{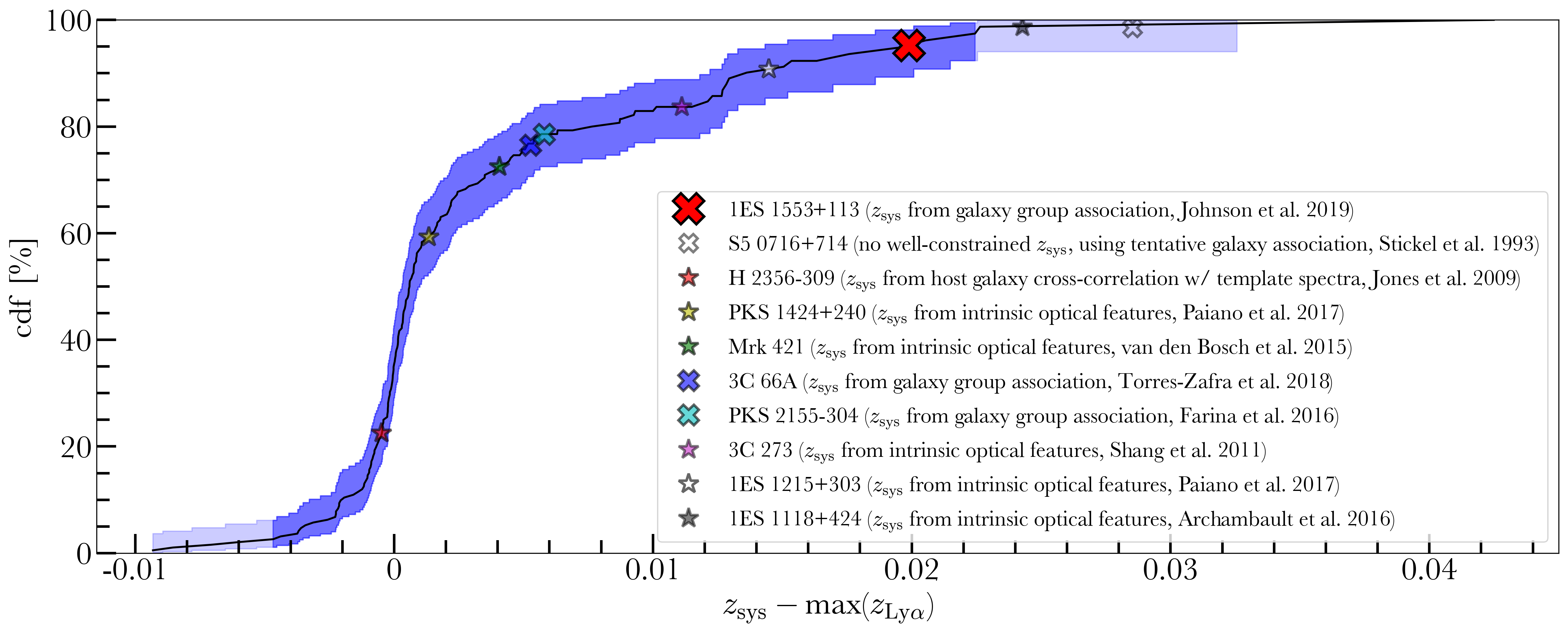}
    \caption{Observed cumulative distribution of the difference between the systemic redshift of an AGN and the highest-redshift \ion{H}{i} Ly$\alpha$ absorption line observed toward it, $\Delta z = z_{\rm{sys}} -$ max($z_{\rm{Ly\alpha}}$), for 192 AGN located at $0.01\,\lesssim\,z\,\lesssim\,0.45$ (see Section~\ref{subsec:forest}, Table~\ref{tab:data}). This distribution thus characterizes the intrinsic scatter in the gap between low-redshift AGN/QSOs and the edge of their Ly$\alpha$ forest. The 95\% confidence interval is highlighted in darker color, which can be used as a robust redshift estimator when combined with an object's UV-detected max($z_{\rm{Ly\alpha}}$) [see equation (1)]. The 10 blazars to which we apply our redshift constraint technique are shown by various symbols according to the legend (see Sections~\ref{subsec:1ESredshift},~\ref{sec:application} and Table~\ref{tab:blazars}).}
    \label{fig:redshift}
\end{figure*}

Finally, we computed the difference between the AGN systemic redshift and its highest-redshift \ion{H}{i} Ly$\alpha$ absorption line for the entire sample: $\Delta z = z_{\rm{sys}} -$ max($z_{\rm{Ly\alpha}}$). The observed cumulative distribution function (CDF) of this dataset is shown in Fig.~\ref{fig:redshift}, which thus depicts the intrinsic scatter in the gap between low-redshift AGN/QSO systemic redshifts and their highest-redshift \ion{H}{i} Ly$\alpha$ absorption line. Negative values in Fig.~\ref{fig:redshift} are AGN with an associated Ly$\alpha$ line beyond their systemic redshift (i.e. max($z_{\rm{Ly\alpha}}$) $> z_{\rm{sys}}$); there are six AGN in our sample with a max($z_{\rm{Ly\alpha}}$) offset by $\Delta v\,<$ -1,000 km s$^{-1}$ (i.e. $\Delta z\,\lesssim\,-0.004$), which could arise from unusually large redshift uncertainty, uncertain line identification, or from interesting, possibly inflowing, gas motions (see Table~\ref{tab:data}).

\begin{table*}
    \centering
    \caption{Low-redshift AGN/QSOs with high-quality \textit{HST}/COS FUV data used for the redshift constraint technique}
    \begin{threeparttable}
    \label{tab:data}
    \begin{tabular}{lllccc}
    \toprule
    \addlinespace
        Name\tnote{a} & $z_{\rm sys}$\tnote{b} & max($z_{\rm{Ly\alpha}}$)\tnote{c} & $z_{\rm source}$\tnote{d} & \textit{HST} Proposal ID &  Measurement flag\tnote{e}\\
    \addlinespace
    \hline
    HE0153-4520 & 0.451 & 0.4284 & VCV & 11541 & 1 \\
    PG0003+158 & 0.4509 & 0.4466 & \citet[][]{Hintzen84} & 12038 & 1 \\
    SDSSJ080359.23+433258.4 & 0.4489 & 0.4481 & SDSS & 11598 & 1 \\
    TON236 & 0.4473 & 0.4480 & SDSS & 12038 & 1 \\
    SDSSJ161649.42+415416.3 & 0.4412 & 0.4352 & SDSS & 11598 & 0\\ 
    \ldots & \ldots & \ldots & \ldots & \ldots & \ldots \\
    \bottomrule
	\end{tabular}
\begin{tablenotes}\footnotesize
\item[]Table~\ref{tab:data} is available online in its entirety in machine-readable format.
\item[a] Object ID, from HSLA (see Section~\ref{subsec:forest})
\item[b] Systemic redshift of the AGN (see Section~\ref{subsec:forest})
\item[c] Highest-redshift Ly$\alpha$ line with $\log N_{\rm{\ion{H}{i}}}/{\rm cm^{-2}} > 12.6$ detected in FUV spectrum of AGN (see Section~\ref{subsec:forest}). For three of the objects that have max($z_{\rm{Ly\alpha}}$)$\,> z_{\rm sys}$ (HB89-202-765, SDSS-J104741.75+151332.2, and SDSSJ213357.89-071217.3), the max($z_{\rm{Ly\alpha}}$) measurement is somewhat uncertain and may be lower, but this does not change our results.
\item[d] Source of $z_{\rm sys}$ (see Section~\ref{subsec:forest})
\item[e] `1' indicates a direct measurement of max($z_{\rm{Ly\alpha}}$); `0' indicates the max($z_{\rm{Ly\alpha}}$) is a right-censored limit due to spectral quality issues (see Section~\ref{subsec:forest})
\end{tablenotes}
\end{threeparttable}
\end{table*}

The 95\% confidence interval (CI) of this distribution can be used as a robust statistical constraint on the redshift, $z$, of any low-redshift AGN/QSO, when combined with the position of its UV-detected max($z_{\rm{Ly\alpha}}$) of $\log N_{\rm{\ion{H}{i}}}/{\rm cm^{-2}} > 12.6$:
\begin{equation}
z_{\rm 95\%\,CI} = 
[{\rm max}(z_{\rm{Ly\alpha}})\,-\,0.0047, 
{\rm max}(z_{\rm{Ly\alpha}})\,+\,0.0224]. 
\end{equation}
The median of the distribution can also be used as an empirical redshift estimate, with two-sided error bars corresponding to the $1\sigma$ error (i.e. 68\% CI):
\begin{equation}
z_{\rm med} = {\rm max}(z_{\rm{Ly\alpha}}) + 
0.0006^{+0.0109}_{-0.0015}.
\end{equation}

In creating the observed cumulative distribution of $\Delta z = z_{\rm{sys}} -$ max($z_{\rm{Ly\alpha}}$) shown in Fig.~\ref{fig:redshift}, we employed standard survival statistics to include 21 AGN ($\approx$11 per cent of the sample) for which spectral quality issues enable only limits on max($z_{\rm{Ly\alpha}}$) by treating the point where the spectrum S/N falls below 10 or where contamination begins as a right-censored data point on $\Delta z$ (see last column of Table~\ref{tab:data}). Of these 21 AGN, six (6) have max($z_{\rm{Ly\alpha}}$) with spectral S/N per resel $<$ 10 (i.e. the spectrum falls below S/N $\approx$ 10 between max($z_{\rm{Ly\alpha}}$) and $z_{\rm{sys}}$), five (5) have a 3$\sigma$ detection limit slightly worse than our NUV spectrum of 1ES 1553+113\footnotemark[4], and 10 AGN spectra contain contamination from geocoronal emission or poor/missing data between max($z_{\rm{Ly\alpha}}$) and $z_{\rm{sys}}$. The distribution in Fig.~\ref{fig:redshift} is thus a Kaplan-Meier curve, computed using \textsc{lifelines} \citep[][]{Lifelines}, which is a non-parametric estimator of the CDF \citep[see, e.g.,][]{Isobe86}.

\subsection{Redshift of 1ES 1553+113 \& Implications for the WHIM}
\label{subsec:1ESredshift}

In Section~\ref{sec:observations}, we verified the redshift lower limit of 1ES 1553+113 from its highest-redshift Ly$\alpha$ line; however, confirmation of its membership with a galaxy group at $z = 0.433$ requires greater certainty in the inferred upper limit for the blazar's redshift based on the edge of the Ly$\alpha$ forest observed toward it (see Section~\ref{subsec:comparison}).

We apply our redshift constraint technique obtained in Section~\ref{subsec:forest} to 1ES 1553+113 using its confirmed highest-redshift Ly$\alpha$ line [i.e. equation (1) with max($z_{\rm{Ly\alpha}}$) = 0.4131] to infer a new 95\% confidence interval for its redshift of $z = 0.408\textrm{--}0.436$. The red X symbol in Fig.~\ref{fig:redshift} at the $\approx$95$^{\rm th}$ percentile shows the position of 1ES 1553+113 within the inferred 95\% confidence interval.

Our empirical redshift constraint for 1ES 1553+113 improves the certainty in its inferred redshift upper limit (see Section~\ref{subsec:comparison}) and confirms the blazar's membership with a $z = 0.433$ galaxy group identified in the deep spectroscopic redshift survey performed by \citet[][]{PaperI}. The stronger candidate \ion{O}{vii} absorber detected toward 1ES 1553+113 at $z = 0.4339$ (\citealt[][]{Nicastro18a}) most likely arises from the CGM or IGrM of the blazar's host environment and cannot be used for the census of warm-hot baryons in the sheets/filaments of the cosmic web \citep[see absorber analysis in][]{PaperI}. Although the available sample of extragalactic X-ray absorption spectra is small, this suggests that the WHIM is not sufficiently metal enriched or is too low density to be detected in \ion{O}{vii} or \ion{O}{viii} with current X-ray facilities (see, e.g., \citealt[][]{Bregman19, Nicastro21}), except possibly via spectral stacking (e.g., \citealt[][]{Kovacs19, Ahoranta20}, also see \citealt[][]{Ahoranta21}). In any case, further X-ray spectroscopy of AGN and blazars with well constrained redshifts and complementary UV spectra and deep galaxy surveys are necessary in order to improve our understanding of the physical properties of the WHIM and its relationship to galaxy feedback and evolution.

\subsection{Systematics of the redshift constraint technique}
\label{subsec:systematics}

In this subsection, we discuss and quantify systematic uncertainties in our blazar redshift constraint technique that may arise from differences between blazars and typical AGN/QSOs due to local environment (Section~\ref{subsubsec:env}) and the proximity effect (Section~\ref{subsubsec:proximity}). We investigate systematic effects that could cause blazars to have an altered distribution of $\Delta z = z_{\rm{sys}} -$ max($z_{\rm{Ly\alpha}}$) compared to the one for AGN/QSOs shown in Fig.~\ref{fig:redshift}. In particular, given that most of the AGN in our sample have a max($z_{\rm{Ly\alpha}}$) that would be considered associated or intrinsic (see Fig.~\ref{fig:redshift}), we focus on quantifying any potential difference in the incidence of associated absorption in the spectra of blazars (e.g., \citealt[][]{Ghisellini11, Chand12, Mishra19, Massaro20}) compared to AGN (e.g., \citealt[][]{Muzahid13, Joshi13, Ganguly13, Chen18}).

\subsubsection{The impact of local environment}
\label{subsubsec:env}

We first consider the impact that known differences in the local environments of blazars and AGN could have on their frequency of associated \ion{H}{i} absorbers. To do this, we examined separately the distributions of $\Delta z = z_{\rm{sys}} -$ max($z_{\rm{Ly\alpha}}$) for radio-loud and radio-quiet AGN, as radio-loud AGN reside in more overdense environments \citep[e.g.,][]{Retana-Montenegro17}, similar to those typical of blazars \citep[e.g.,][]{Allevato14}. Indeed, BL Lac objects reside in galaxies that are on average more massive than AGN/QSO host galaxies (see, e.g., \citealt[][]{Urry00} and the review by \citealt[][]{Falomo14}), which could possibly result in a different distribution of associated absorbers.

We cross-correlated our AGN sample with the Catalog of Quasar Properties from SDSS DR7 \citep[][]{Shen11} to determine which objects have radio measurements from the FIRST survey \citep[][]{Helfand15}. Adopting the definition for radio-loudness from \citet[][]{Kellermann89}, we identified 16 radio-loud AGN and 60 confirmed radio-quiet AGN in our sample. We find that the radio-loud and radio-quiet AGN have very similar distributions of $\Delta z = z_{\rm{sys}} - $max($z_{\rm{Ly\alpha}}$). Performing a two-sample Kolmogorov--Smirnov (K--S) test yields a statistic of 0.167 with p-value of 0.82, although it is important to note that the sample size of radio AGN is likely too small to detect any significant difference between the $\Delta z$ distributions. The median $\Delta z$ of the radio-quiet AGN is 0.001 compared to the median of 0.00002 for the radio-loud AGN. The difference between these median $\Delta z$ values is $\approx 5 \times$ less than the scatter of the distributions. This confirms that the difference in local environment between blazars and AGN does not have a significant effect on their observed distributions of $\Delta z$, and so does not contribute to systematics of the redshift constraint technique.

\subsubsection{The impact of the proximity effect}
\label{subsubsec:proximity}

While the presence of associated absorbers in AGN and blazar spectra is expected to tighten the $\Delta z$ distribution, the proximity effect \citep[e.g.,][]{Bajtlik88, Kulkarni93, Bechtold94, Pascarelle01, Calverley11} due to enhanced ionization radiation near the AGN can decrease the incidence of absorbers which would broaden the distribution. Moreover, because blazars are more highly variable than typical AGN, the proximity effect around them may be different due to increased importance of non-equilibrium effects. To quantify the potential impact of the proximity effect on Ly$\alpha$ absorption inferred redshifts, we first searched for a trend between AGN luminosity and $\Delta z$ and then placed a conservative upper limit on the potential impact of the proximity effect by examining a spectral region where it is significant.

To determine whether the proximity effect can bias Ly$\alpha$ inferred redshifts, we calculated the generalized Kendall rank correlation coefficient between the AGN UV luminosity \citep[][]{Bianchi14} and $\Delta z = z_{\rm{sys}} -$ max($z_{\rm{Ly\alpha}}$) and found $\tau=+0.13$ with a marginal significance of $0.04$. Moreover, this weak correlation is largely driven by the three lowest luminosity AGN in the sample. This experiment demonstrates that the proximity effect does not contribute significantly to the systematic uncertainty in Ly$\alpha$ inferred redshifts for non-blazar AGN.

Blazars are significantly more variable than typical AGN, which may result in a difference in proximity effect caused by non-equilibrium effects. To ensure that our Ly$\alpha$ inferred blazar redshifts are robust, we placed a conservative upper limit on systematic uncertainty resulting from the proximity effect by remeasuring the $\Delta z$ distribution after excluding the spectral region that corresponds to \ion{H}{i} Ly$\alpha$ velocities within 1500 \kms\ of the AGN systemic. By identifying only max($z_{\rm{Ly\alpha}}$) that are at $\Delta v > $1500 \kms\, from each AGN, we maximize the spread in the $\Delta z$ distribution because associated absorbers are less common in this velocity range while the proximity effect remains significant \citep[e.g.][]{Tripp08, Fox08}.

The S/N available in the archival COS AGN spectra are lower at $|\Delta v| > $1500 \kms\ from the AGN systemic velocity because of the weaker Ly$\alpha$ emission away from the line peak. Consequently, when making the `intervening only' $\Delta z$ distribution, we adopted a higher threshold in column density of $\log N_{\rm{\ion{H}{i}}}/{\rm cm^{-2}} >$ 13 to ensure high completeness rates. Finally, we shifted the resulting $\Delta z$ distribution by $-0.006$ in redshift, which corresponds to $\Delta v = -1500$ \kms\ at the median systemic redshift of our AGN sample. This ensures that the resulting cumulative distribution can be used as an approximately unbiased estimator of AGN and blazar redshifts.

The resulting `intervening only' distribution is shown in orange in Fig.~\ref{fig:systematics}. If this curve is adopted to infer redshifts, the resulting 95\% confidence interval and median would be: 
\begin{equation}
z_{\rm 95\%\,CI} = 
[{\rm max}(z_{\rm{Ly\alpha}})\,-\,0.0003, 
{\rm max}(z_{\rm{Ly\alpha}})\,+\,0.0370]
\end{equation}
and
\begin{equation}
z_{\rm med} = {\rm max}(z_{\rm{Ly\alpha}}) + 
0.0080^{+0.0138}_{-0.0052}.
\end{equation}

The resulting redshift constraints (i.e. equations (1-2) and (3-4)) differ in the median by less than the $1\sigma$ uncertainty given in equation (2). Therefore, we have shown that the edge of the Ly$\alpha$ forest can be used to set a robust statistical constraint on the redshift of AGN and blazars with a statistical $1\sigma$ uncertainty of $\approx 0.01$ and with sub-dominant systematic uncertainty coming from any differences in the frequency of associated absorption and/or the proximity effect.

\subsubsection{Analytical comparison}
\label{subsubsec:analyticalcomparison}

To provide an analytical comparison to the observed CDFs of $\Delta z = z_{\rm{sys}} -$ max($z_{\rm{Ly\alpha}}$) shown in Fig.~\ref{fig:systematics}, we use Poisson statistics to calculate the probability of observing no Ly$\alpha$ absorption lines as a function of $\Delta z$ when expecting to observe the mean number of intervening \ion{H}{i} systems per unit redshift, $\langle \frac{d \mathcal{N}}{dz} \rangle$ (also see \citealt[][]{Blades85, Furniss13a}). We adopt the redshift-dependent $\frac{d \mathcal{N}}{dz}$ measurement from \citet[][]{Danforth16}, expressed as $\frac{d \mathcal{N (>\rm{N}, z)}}{d z} = C_0 (1+z)^{\gamma}$, where C$_0 = 91 \pm 1$ and $\gamma = 1.24 \pm 0.04$ (see their fig.~13), and we use this to calculate the probability mass function as $P(z) = \frac{\lambda^k e^{-\lambda}}{k!}$, where $k=0$ and $\lambda = \langle \frac{d \mathcal{N}}{dz} \rangle \times$ ($z -$ max($z_{\rm{Ly\alpha}}$)). The resulting expected Poisson-based CDF is plotted in Fig.~\ref{fig:systematics} for max($z_{\rm{Ly\alpha}}$)$= 0.4$.

We see that the Poisson-based CDF is qualitatively similar to the two observed CDFs, falling in between them. The blue cumulative distribution appears to the left of the expected CDF because it is created using mostly associated max($z_{\rm{Ly\alpha}}$), which drives it to lower, and even negative, values of $\Delta z$. The orange cumulative distribution appears more similar in shape to the expected CDF, but it is nonetheless driven to slightly larger values of $\Delta z$. Because the two distributions are obtained using the same minimum detected column density for intervening \ion{H}{i}, the observed difference between them is likely due to the proximity effect near AGN (see Section~\ref{subsubsec:proximity}).

\subsection{The impact (and benefit) of IGM evolution}
\label{subsec:evolution}
The number of Ly$\alpha$ absorbers per unit redshift evolves both because of the expansion of the Universe and physical evolution of Ly$\alpha$ forest clouds \citep[e.g.,][]{Rauch98}. Consequently, we expect both the median and scatter in the gap between AGN systemic redshifts and the highest redshift Ly$\alpha$ absorber in their spectra to evolve with redshift. To estimate the significance of this evolution, we re-evaluated the  CDFs for all (see Figure \ref{fig:redshift}) and intervening (see Figure \ref{fig:systematics}) absorbers after splitting the AGN sample by redshift at $z=0.2$.

Somewhat surprisingly, when using both associated and intervening absorbers, the CDF does not change significantly. This suggests that the frequency of associated absorbers does not evolve over the relevant redshift range. For this reason, we adopt equations (1) and (2) as our preferred redshift estimators and use them even at a somewhat higher redshift in Section~\ref{subsec:consistent}. On the other hand, the change in the CDF formed from intervening-only absorbers with $\Delta v>1500$ \kms\ is more significant. In particular, for AGN with $z<0.2$, we find $dz_{\rm med}=0.009$ and an upward going $1\sigma$ uncertainty of $+0.022$. For AGN at $z=0.2-0.45$, these decrease to $dz_{\rm med}=0.007$ and a $1\sigma$ uncertainty of $+0.019$.

At higher redshifts of $z>0.4$, we expect the trend of decreasing $dz_{\rm med}$ and smaller uncertainty to continue due to the thickening of the Ly$\alpha$ forest. To estimate the significance of this effect at $z>0.4$, we adopt the redshift evolution estimate of $\frac{dN}{dz} \propto (1 + z)^{1.24}$ from \citet[][]{Danforth16} and compute expected CDFs assuming Poisson statistics as described in Section~\ref{subsubsec:analyticalcomparison}. This results in an expected decrease of $\approx 0.002$ in $dz_{\rm med}$ between $z=0.4$ and $z=0.8$. Similarly, the expected confidence intervals are $\approx 25$ per cent tighter at $z=0.8$ relative to $z=0.4$. This further motivates the expansion of our technique to higher redshifts ($0.5 \lesssim z \lesssim 1$) through new COS NUV spectra of AGN, QSOs, and blazars, especially because the current NUV archival sample is not large enough to produce an empirical CDF at $z>0.45$.

\subsection{Comparison to previous methods}
\label{subsec:comparison}

\begin{figure*}
    \includegraphics[width=\linewidth]{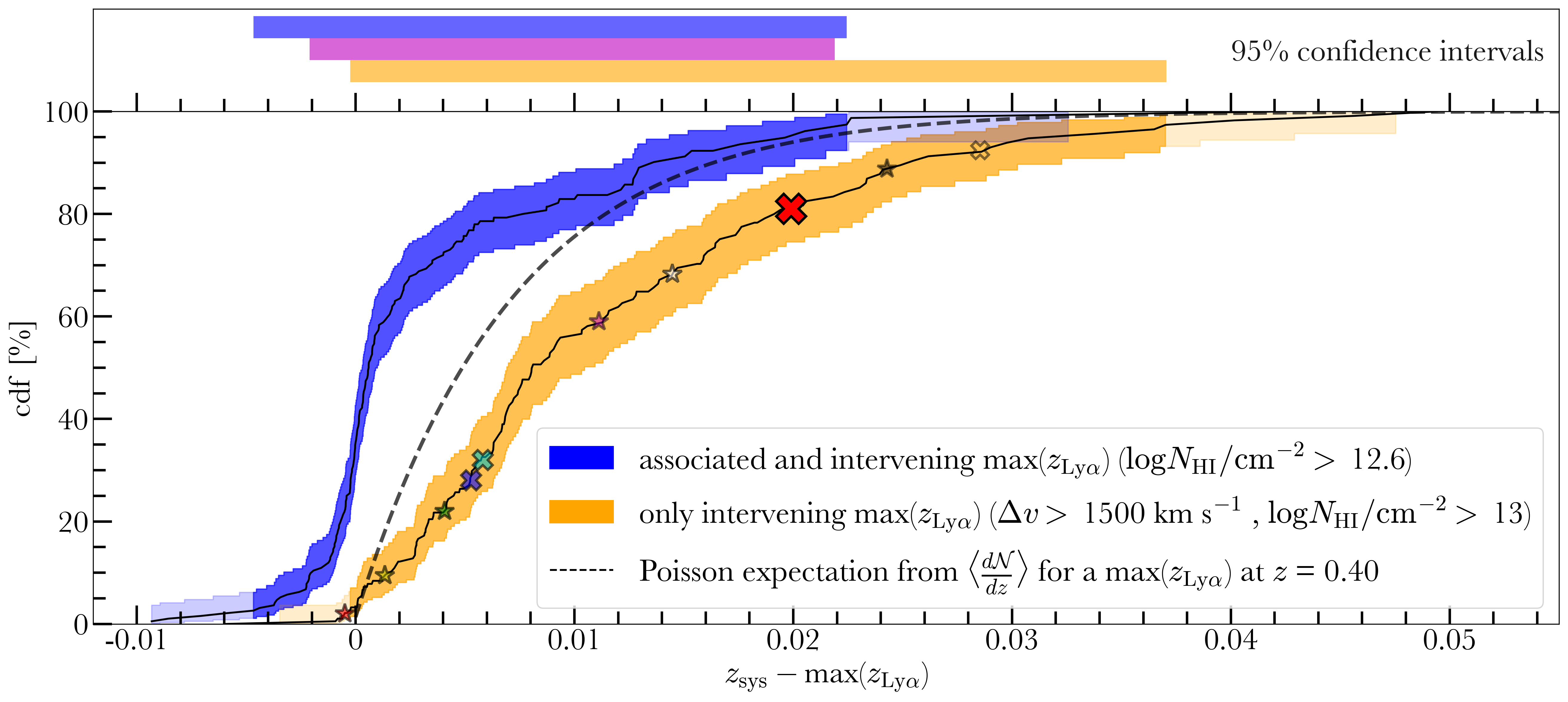}
    \caption{Cumulative distributions of $\Delta z = z_{\rm{sys}} -$ max($z_{\rm{Ly\alpha}}$) as described in Fig.~\ref{fig:redshift}. The blue distribution is the same as in Fig.~\ref{fig:redshift}, and the orange distribution represents a conservative estimate of the systematic uncertainty in the redshift constraint due to the proximity effect (see Section~\ref{subsubsec:proximity}). The orange curve is obtained using only the highest-redshift intervening Ly$\alpha$ lines with $\log N_{\rm{\ion{H}{i}}}/{\rm cm^{-2}} > 13$ (see definition in the main text), shifted to $\Delta z = 0$ to be used as an approximately unbiased redshift constraint in the case that blazars have a lower incidence of associated Ly$\alpha$ (see Section~\ref{subsec:systematics}). The 10 blazars to which we apply our redshift constraint technique are shown on the orange curve by various symbols according to the legend in Fig.~\ref{fig:redshift} to represent their alternative redshift constraints than the ones adopted in Section~\ref{sec:application} (see Section~\ref{subsec:systematics}, Table~\ref{tab:blazars}). The black dashed curve shows the expectation from Poisson statistics calculated based on the probability of observing no \ion{H}{i} Ly$\alpha$ lines as a function of $\Delta z$ assuming the $\frac{d \mathcal{N}}{dz}$ measurement from \citealt[][]{Danforth16} at $z = 0.40$ (see Section~\ref{subsubsec:analyticalcomparison}). The top panel compares our empirical redshift constraints to the constraint from \citet[][]{PaperI}, which shares the same methodology: blue -- ours (mostly associated and some intervening Ly$\alpha$, Section~\ref{subsec:forest}); orange -- ours (only intervening Ly$\alpha$, see Section~\ref{subsubsec:proximity}); purple -- \citet[][]{PaperI} (based on a reference AGN sample $\approx 1/3$ the size of ours, mostly associated and some intervening Ly$\alpha$, see Section~\ref{subsec:comparison}). We also note that our methodology results in a redshift constraint that is $\approx 5 \times$ tighter than previous methods used to constrain blazar redshifts from their \ion{H}{i} Ly$\alpha$ forest (\citealt[][]{Danforth10}, see Section~\ref{subsec:comparison}). 
    }
    \label{fig:systematics}
\end{figure*}

An updated redshift constraint for 1ES 1553+113 of $z = 0.411\textrm{--}0.435$ was obtained by \citet[][]{PaperI} using the method ours is based on (i.e. applying the observed AGN/QSO distribution of $\Delta z = z_{\rm{sys}} -$ max($z_{\rm{Ly\alpha}}$) to blazars), which was used to infer the blazar's membership with a galaxy group they identified at $z = 0.433$. To measure the observed distribution of $\Delta z = z_{\rm{sys}} -$ max($z_{\rm{Ly\alpha}}$), they used the AGN sample from \citet[][]{Danforth16}\footnotemark[3], whereas we utilize the entirety of the COS archive available through HSLA to construct an AGN sample of high-S/N Ly$\alpha$ forest sightlines that is $\approx 3 \times$ larger. Consequently, our observed CDF of $\Delta z = z_{\rm{sys}} -$ max($z_{\rm{Ly\alpha}}$) is better populated at the wings, and so the resulting 95\% confidence interval (i.e. our redshift constraint) is more robust. Furthermore, we have shown that the precision of our redshift constraint is not significantly affected if we assume blazars to have a lower incidence of associated Ly$\alpha$ than most AGN [see Fig.~\ref{fig:systematics} and equations (1) and (3)].

Preceding this, one of the first instances of the Ly$\alpha$ forest being used to constrain the redshift of a blazar was by \citet[][]{Danforth10}, who utilized the Ly$\alpha$ and Ly$\beta$ forests in the FUV spectrum of 1ES 1553+113 to infer a redshift constraint of $0.433 < z_{\rm{sys}} \lesssim 0.58$ (at a 1$\sigma$ confidence limit). The redshift lower limit was set by the highest-redshift Ly$\alpha$ line; however, this was based on a misidentification of \ion{Ni}{ii} as \ion{H}{i} Ly$\alpha$ and was later corrected by \citet[][]{Danforth16} and \citet[][]{PaperI}. When using the correct line identification (see Fig.~\ref{fig:spectrum}), the redshift constraint becomes $0.413 < z_{\rm{sys}} \lesssim 0.56$. The redshift upper limit was obtained by applying a K--S test to differentiate between the expected and observed (non-detection) CDFs of \ion{H}{i} absorbers ($\langle \frac{d \mathcal{N}}{dz dN} \rangle$) over the range $0.4 < z < 0.75$ (also see \citealt[][]{Danforth13}). This method requires binning the observed absorbers by redshift (with a chosen bin size of $\Delta z = 0.025$). Moreover, the K--S test is not particularly sensitive to differences near the ends (i.e. wings) of the CDFs \citep[see, e.g., Section 3.1 of][]{FeigelsonBabu12}. By contrast, our technique does not require binning and uses a full, empirical CDF resulting in a factor of $\approx 5\times$ improvement in redshift constraining power.

\section{Application of redshift constraint technique to other blazars}
\label{sec:application}

As discussed in Section~\ref{intro}, blazars are useful cosmological tools primarily because of their brightness in the UV, X-ray, and $\gamma$-ray. However, many blazars -- so-called BL Lac objects such as 1ES 1553+113 -- are so bright, relative to their host galaxy and any nuclear line emission, that intrinsic spectral features are nearly or completely undetectable, making it difficult to determine their redshifts. Despite deep observations across the electromagnetic spectrum, many such quasi-featureless blazars have no direct spectroscopic redshift measurement, and so indirect methods are required to constrain their redshifts instead, such as by associating the blazar with a nearby galaxy group \citep[e.g.,][]{Farina16, Rovero16, Torres-Zafra18, PaperI}.

Given the agreement between our redshift constraint for 1ES 1553+113 and its inferred galaxy group association (see Sections~\ref{subsec:forest},~\ref{subsec:1ESredshift}), we further test the utility of our redshift constraint technique by applying it to a number of other well-known blazars that have difficult-to-measure redshifts. We searched the literature to construct a sample of nine (9) additional bright blazars that have archival \textit{HST} UV spectra, most of which are key targets for future studies of the WHIM \citep[][]{Bregman15} and for constraints on the EBL \citep{BiteauWilliams}. We use our updated Ly$\alpha$ forest based redshift estimation technique\footnotemark
\footnotetext{The following applications of our redshift constraint technique to blazars use equation (1) (i.e. the 95\% confidence interval obtained in Section~\ref{subsec:forest}). This application inherently assumes that blazars and AGN have a similar incidence of associated Ly$\alpha$ because the majority of the max($z_{\rm{Ly\alpha}}$) used to obtain our redshift constraint [i.e., equation (1)] is `associated' (see Section~\ref{subsec:systematics}). However, this assumption may not be valid given possible differences between blazars and AGN/QSOs. Because of this, in Section~\ref{subsubsec:proximity} we explored the appropriate redshift constraint resulting from our technique if blazars and AGN/QSOs were to have different observed incidences of associated \ion{H}{i} Ly$\alpha$, finding that the inferred redshift lower and upper limits would be expected to increase by only $\approx0.004$ and $\approx0.015$, respectively [see Fig.~\ref{fig:systematics} and equations (1) and (3)].} to establish more certainty in (Section~\ref{subsec:consistent}) and even improve (Section~\ref{subsec:improves}) the redshift estimates of these blazars (see Table~\ref{tab:blazars}). In Fig.~\ref{fig:redshift}, we show the positions of the blazars in relation to the adopted redshift constraint, and, in Fig.~\ref{fig:systematics}, we show the blazars in relation to the alternative redshift constraint that corresponds to blazars having a lower incidence of associated \ion{H}{i} Ly$\alpha$ (see Section~\ref{subsubsec:proximity}).

\subsection{Constraint is consistent with blazar's archival redshift}
\label{subsec:consistent}

\begin{enumerate}
\item \textbf{Mrk 421}
As one of the nearest blazars to the MW, Mrk 421 is well-studied for its extreme brightness and variability (e.g., \citealt[][]{Ulrich97, Abdo11}). Despite multiple X-ray observations, there are so far no confident detections of the hot WHIM toward Mrk 421 in \ion{O}{vii} or \ion{O}{viii} absorption (\citealt[][]{Nicastro05b, Kaastra06, Rasmussen07, Danforth11, Yao12, Nicastro14}).

The redshift of Mrk 421 has been measured from intrinsic optical features, first to be $z = 0.0300\,\pm\,0.0001$ (\citealt[][]{deVaucouleurs91}) and later updated to be $z = 0.0293\,\pm\,0.0001$ (\citealt[][]{vandenBosch15}). We utilize the archival COS G130M+G160M FUV spectra of Mrk 421 to identify its highest-redshift Ly$\alpha$ absorption line. Based on its max($z_{\rm{Ly\alpha}}$) at $z = 0.02523$, we infer a redshift constraint for Mrk 421 using our technique [i.e. equation (1)] of $z_{\rm sys}$ = 0.021--0.048, which is consistent with its systemic redshift ($\approx$73$^{\rm rd}$ percentile, see Fig.~\ref{fig:redshift}).

\item \textbf{PKS 2155-304}
This BL Lac-type blazar is one of the highest priority sightlines for future X-ray absorption studies of the hot WHIM, given its considerable X-ray flux and comparatively long redshift pathlength \citep[][]{Bregman15}. Tentative detections of intervening \ion{O}{viii} have been reported toward PKS 2155-304 \citep[][]{Fang02, Fang07} but have not been confirmed by follow-up observations \citep[][]{Rasmussen03, Yao10, Nevalainen19}, marking the importance of future X-ray missions to clarify the possible hot gas along this sightline.

The redshift of PKS 2155-304 has been indirectly constrained through association with a nearby galaxy overdensity at $z = 0.116$ (\citealt[][]{Falomo91b, Falomo93}), later confirmed and refined to be located at $z = 0.11610\,\pm\,0.00006$ (\citealt[][]{Farina16}). We utilize the archival COS G130M FUV spectrum of PKS 2155-304 to identify its highest-redshift Ly$\alpha$ absorption line, which we confirm with the line identifications of \citet[][]{Danforth16}\footnotemark[3]. Based on its max($z_{\rm{Ly\alpha}}$) at $z = 0.11029$, our inferred redshift constraint for PKS 2155-304 is $z_{\rm sys}$ = 0.106--0.133, which is consistent with membership in the galaxy group at $z = 0.116$ ($\approx$79$^{\rm th}$ percentile, see Fig.~\ref{fig:redshift}).

\item \textbf{1ES 1215+303}
This TeV-detected, $\gamma$-ray blazar is of particular interest to the high-energy astrophysics community (e.g., \citealt[][]{Aleksic12, Valverde20}) and is considered a priority sightline for future X-ray studies of the WHIM (\citealt[][]{Bregman15}). 

Its redshift has been determined through detection of intrinsic optical features ([\ion{O}{ii}] and [\ion{O}{iii}]) at $z \approx 0.131$ (\citealt[][]{Paiano17}), validating the previous estimate of $z = 0.130$ reported by \citet[][]{Bade98}. This redshift measurement was further verified by detection of the blazar's broad Ly$\alpha$ emission line in the FUV (\citealt[][]{Furniss19}). We utilize the archival COS G130M+G160M FUV spectra of 1ES 1215+303 to identify its highest-redshift Ly$\alpha$ absorption line. Based on its max($z_{\rm{Ly\alpha}}$) at $z = 0.11652$, our inferred redshift constraint for 1ES 1215+303 is $z_{\rm sys}$ = 0.112--0.139, which is consistent with its previously measured systemic redshift of $z = 0.131$ ($\approx$91$^{\rm st}$ percentile, see Fig.~\ref{fig:redshift}).

\item \textbf{3C 273}
Observations of this bright FSRQ blazar \citep[e.g.,][]{Healey07} have yielded some of the first reported detections of the highly-ionized IGM: The FUV spectrum of 3C 273 contains intervening \ion{O}{vi} absorption at $z = 0.120$ \citep[][]{Sembach01, Danforth08}, and a recent X-ray follow-up study of 3C 273 reported the tentative detection of the hot WHIM through \ion{O}{viii} and \ion{Ne}{ix} absorption at $z = 0.090$ (\citealt[][]{Ahoranta20}).

The redshift of 3C 273 is known from intrinsic optical features (\ion{Mg}{ii}, [\ion{O}{iii}], and Balmer lines) found at $z = 0.158$ (\citealt[][]{Schmidt63}), later refined to be $z = 0.1576$ using the [\ion{O}{iii}] line (\citealt[][]{Shang11}). We utilize the archival COS G130M FUV spectrum of 3C 273 to identify its highest-redshift Ly$\alpha$ absorption line, which we confirm with the line identifications of \citet[][]{Danforth16}\footnotemark[3]. Based on its max($z_{\rm{Ly\alpha}}$) at $z = 0.14648$, our inferred redshift constraint for 3C 273 is $z_{\rm sys}$ = 0.142--0.169, which is consistent with its systemic redshift ($\approx$84$^{\rm th}$ percentile, see Fig.~\ref{fig:redshift}).

\item \textbf{H 2356-309}
This BL Lac object is considered a top-20 AGN sightline for absorption studies of the hot WHIM \citep[][]{Bregman15}. So far, no detections of intervening \ion{O}{vii} or \ion{O}{viii} lines have been reported toward H 2356-309, although a transient \ion{O}{viii} absorption feature was found intrinsic to the blazar \citep[][]{Fang11}.

The redshift of H 2356-309 was first measured from stellar absorption features at $z = 0.165$ \citep[][]{Falomo91a} and was later refined to be $z = 0.16539\,\pm\,0.00018$ by cross-correlating the host galaxy spectrum with predefined template spectra \citep[][]{6dF}. Its redshift was further confirmed by the detection of its weak Ly$\alpha$ emission line \citep[][]{Fang14}. We utilize the archival COS G130M FUV spectrum of H 2356-309 to identify its highest-redshift Ly$\alpha$ absorption line, which we confirm with the line identifications of \citet[][]{Danforth16}\footnotemark[3]. Based on its max($z_{\rm{Ly\alpha}}$) at $z = 0.16588$, our inferred redshift constraint for H 2356-309 is $z_{\rm sys}$ = 0.161--0.188, which is consistent with its systemic redshift ($\approx$23$^{\rm rd}$ percentile, see Fig.~\ref{fig:redshift}).

\item \textbf{3C 66A}
This BL Lac object is one of the brightest extragalactic $\gamma$-ray sources \citep[e.g.,][]{Acciari09} and has a nearly featureless spectrum, which has resulted in spurious redshift measurements over time \citep[see][and references therein]{Bramel05}. The lack of a firm spectroscopic redshift measurement for this blazar motivated follow-up observations that revealed its likely association with a galaxy group at $z = 0.340$ \citep[][]{Torres-Zafra18}. We utilize the archival COS G130M+G160M FUV spectra of 3C 66A to identify its highest-redshift Ly$\alpha$ absorption line, which we confirm with the line identifications of \citet[][]{Danforth16}\footnotemark[3]. Based on its max($z_{\rm{Ly\alpha}}$) at $z = 0.33472$, our inferred redshift constraint for 3C 66A is $z_{\rm sys}$ = 0.330--0.357, which is consistent with its galaxy group association ($\approx$77$^{\rm th}$ percentile, see Fig.~\ref{fig:redshift}) and notably improves its previous redshift upper limit inferred using its Ly$\alpha$ forest \citep[][]{Furniss13a}. We note that the updated redshift for 3C 66A of $z = 0.34$ \citep[][]{Torres-Zafra18} can be used to improve studies of the EBL performed toward this blazar \citep[see][]{Dominguez11}.

\item \textbf{PKS 1424+240}
This TeV-detected BL Lac object is one of the highest-redshift known blazars and so may be useful for absorption studies of the hot WHIM and for constraints on the EBL. Its redshift has been indirectly constrained to be $z = 0.6010\,\pm\,0.003$ through association with a nearby galaxy group \citep[][]{Rovero16}, which is consistent with its previous redshift lower limit based on its Ly$\beta$ forest \citep[][]{Furniss13b} and was subsequently refined by the identification of intrinsic emission lines ([\ion{O}{ii}] and [\ion{O}{iii}]) at $z = 0.6047$ \citep[][]{Paiano17}.

We utilize the archival COS G130M+G160M FUV spectra and Space Telescope Imaging Spectrograph (STIS) NUV spectrum (PI: Furniss, PID: 13288) of PKS 1424+240 to identify its highest-redshift Ly$\alpha$ absorption line, which we confirm with the line identifications of \citet[][]{Danforth16}\footnotemark[3]. Based on its max($z_{\rm{Ly\alpha}}$) at $z = 0.60336$, our inferred redshift constraint for PKS 1424+240 is $z_{\rm sys}$ = 0.599--0.626, which is consistent with both its galaxy group association and also the detected intrinsic spectral features ($\approx$7$^{\rm th}$ and $\approx$59$^{\rm th}$ percentiles, respectively, see Fig.~\ref{fig:redshift}). Our redshift constraint for PKS 1424+240 significantly improves its previously inferred redshift upper limits of $z < 0.64$ and $z < 0.75$ based on the blazar's $\gamma$-ray \citep[][]{BiteauWilliams} and FUV \citep[][]{Furniss13b} spectra, respectively. We note that this redshift constraint would be an estimated $\approx$ 15 per cent tighter if we accounted for the expected change in the evolution of \ion{H}{i} systems \citep[][]{Danforth16} over $0.45 < z < 0.6$ (see Section~\ref{subsec:evolution}).
\end{enumerate}

\begin{table*}
    \centering
    \caption{Blazars with archival \textit{HST} UV data to which we apply our redshift constraint technique}
    \begin{threeparttable}
    \label{tab:blazars}
    \begin{tabular}{lllccc}
    \toprule
    \addlinespace
        Name\tnote{a} & $z_{\rm Lit}$\tnote{b} & max($z_{\rm{Ly\alpha}}$)\tnote{c} & $z_{\rm 95\%\,CI}$\tnote{d} & note\tnote{e} & \textit{HST} Proposal ID(s) \\
    \addlinespace
    \hline
    Mrk 421 & 0.0293 & 0.02523 & [0.021,0.048] & consistent & 11520,12025\\
    PKS 2155-304 & 0.1161 & 0.11029 & [0.106,0.133] & consistent & 12038\\
    1ES 1215+303 & 0.131 & 0.11652 & [0.112,0.139] & consistent & 13651\\
    3C 273 & 0.1576 & 0.14648 & [0.142,0.169] & consistent & 12038\\
    H 2356-309 & 0.1654 & 0.16588 & [0.161,0.188] & consistent & 12864\\
    1ES 1118+424 & 0.230 & 0.20572 & [0.201,0.228] & improves & 14772\\
    S5 0716+714 & 0.26 & 0.23146 & [0.227,0.254] & improves & 12025\\
    3C 66A & 0.340 & 0.33472 & [0.330,0.357] & consistent & 12612,12863\\
    1ES 1553+113 & 0.433 & 0.41306 & [0.408,0.436] & consistent & 11520,12025,15835\\
    PKS 1424+240 & 0.6047 & 0.60336 & [0.599,0.626] & consistent & 12612,13288\\
    \bottomrule
	\end{tabular}
\begin{tablenotes}\footnotesize
\item[a] Object ID, from HSLA (see Section~\ref{sec:application})
\item[b] Redshift estimate for blazar from the literature (see references in Section~\ref{sec:application}).
\item[c] Highest-redshift Ly$\alpha$ line detected in UV spectrum of blazar (see Section~\ref{sec:application}). For 1ES 1118+424, additional UV observations are necessary to confirm its max($z_{\rm{Ly\alpha}}$) (see Section~\ref{subsec:improves}).
\item[d] Redshift constraint (95\% confidence interval) for blazar based on the observed edge of the Ly$\alpha$ forest obtained using our technique (see Sections~\ref{subsec:forest},~\ref{subsec:systematics},~\ref{sec:application})
\item[e] `consistent' indicates $z_{\rm 95\%\,CI}$ is consistent with the blazar's systemic redshift (see Section~\ref{subsec:consistent}), `improves' indicates $z_{\rm 95\%\,CI}$ improves the blazar's redshift estimate (see Section~\ref{subsec:improves})
\end{tablenotes}
\end{threeparttable}
\end{table*}

\subsection{Constraint improves blazar's redshift estimate}
\label{subsec:improves}

\begin{enumerate}
\item \textbf{1ES 1118+424}
There is some discrepancy surrounding the redshift of this BL Lac object (which is also known as RBS 0970 or FBQS J112048.0+421212); stellar absorption features (\ion{Ca}{ii}, \ion{Ca}{i}, and \ion{Mg}{i}) were detected at $z = 0.230$ \citep[][]{Archambault16}, but, subsequently, a redshift lower limit of $z > 0.28$ was inferred based on the apparent non-detection of stellar absorption features expected from the host galaxy (see \citealt[][]{Paiano17} and references therein). We utilize the archival COS G130M FUV spectrum of 1ES 1118+424 to identify its highest-redshift Ly$\alpha$ absorption line. Based on its max($z_{\rm{Ly\alpha}}$) at $z = 0.20572$, our inferred redshift constraint for 1ES 1118+424 is $z_{\rm sys}$ = 0.201--0.228, which is possibly in tension with the previous estimate of $z = 0.230$ ($\approx$99$^{\rm th}$ percentile, see Fig.~\ref{fig:redshift}). Given the lack of G160M coverage toward this blazar, additional UV observations are necessary to confirm that there are no further \ion{H}{i} Ly$\alpha$ lines beyond $z = 0.20572$. Additional galaxy redshift surveys in the field would also be useful to further explore association with a host group.

\item \textbf{S5 0716+714}
This bright and highly-variable \citep[][]{Bach07} BL Lac object is considered a top-50 AGN sightline for absorption studies of the hot WHIM \citep[][]{Bregman15}; however, it has no direct spectroscopic redshift measurement despite numerous attempts to detect its host galaxy \citep[][]{Biermann81, Stickel93, Scarpa00, Rector01, Pursimo02, Bychkova06, Finke08, Shaw13, Paiano17}. The edge of the Ly$\alpha$ forest toward this blazar has been used to set a statistical constraint on its redshift of $0.2315 < z < 0.322$ (95\% confidence interval, \citealt[][]{Danforth13}, see description of method in Section~\ref{subsec:comparison}), which was noted to be consistent with its previous redshift estimate of $z = 0.31\,\pm\,0.08$ based on the tentative photometric detection of its host galaxy \citep[][]{Nilsson08}. Alternatively, S5 0716+714 may be associated with a nearby pair of galaxies at $z = 0.26$ \citep[][]{Stickel93, Danforth13}. Studies of the EBL-corrected $\gamma$-ray spectrum of S5 0716+714 have shown both redshift estimates to be possible \citep[][]{Anderhub09, MAGIC18}, demonstrating the need for a more confident redshift constraint for this blazar.

We utilize the archival COS G130M+G160M FUV spectra of S5 0716+714 to identify its highest-redshift Ly$\alpha$ absorption line, which we confirm with the line identifications of \citet[][]{Danforth16}\footnotemark[3]. Based on its max($z_{\rm{Ly\alpha}}$) at $z = 0.23146$, our inferred redshift constraint for S5 0716+714 using equation (1) is $z_{\rm sys}$ = 0.227--0.254, which provides a strong redshift constraint for this blazar. Our improved upper limit for the redshift of S5 0716+714 suggests that it may indeed be a member of the galaxy group at $z = 0.26$ ($\approx$99$^{\rm th}$ percentile, see Fig.~\ref{fig:redshift}), especially if we assume blazars to have a lower average incidence of associated Ly$\alpha$ (see Section~\ref{subsec:systematics}, Fig.~\ref{fig:systematics}). However, our redshift constraint for S5 0716+714 does leave open the possibility that it is located at even lower redshift, and so further deep galaxy surveys toward this blazar are necessary to confirm its redshift.
\end{enumerate}

In summary, out of the ten (10) blazars we have considered in this work (Table~\ref{tab:blazars}), our redshift constraint technique improves the redshift estimates of two (2) blazars (1ES 1118+424 and S5 0716+714, see Section~\ref{subsec:improves}) and is consistent with the systemic redshifts of the other eight (8) blazars (see Sections~\ref{subsec:1ESredshift},~\ref{subsec:consistent}). These blazars, among others, are key targets both for future X-ray missions that aim to study the WHIM \citep[e.g.,][]{Bregman15} and also for constraints on the EBL in the $\gamma$-ray \citep[e.g.,][]{BiteauWilliams}. Indeed, the most X-ray bright blazars are BL Lac objects, whereas FSRQ-type blazars have systematically lower X-ray flux \citep[e.g.,][]{Massaro15, Mao17, Sharma21}, further establishing the importance of accurate redshift constraints for featureless blazars/BL Lac objects. We have thus demonstrated the utility of the edge of the observed \ion{H}{i} Ly$\alpha$ forest as an independent redshift constraint for blazars, which may be crucial for interpreting future studies of the hot WHIM, as we have indeed shown to be the case for 1ES 1553+113 (see Section~\ref{subsec:1ESredshift}).

\section{Conclusions}
\label{sec:conclusions}

In this paper, we have utilized the edge of the observed \ion{H}{i} Ly$\alpha$ forest as a robust statistical tool to improve the precision of indirect redshift constraints for featureless blazars. We developed a technique to constrain the redshift of an AGN or blazar -- with a $1\sigma$ uncertainty of $\approx 0.01$ -- using only the position of its UV-detected, highest-redshift \ion{H}{i} Ly$\alpha$ absorption line. Our technique is based on a large sample of 192 AGN/QSOs at $0.01\,\lesssim\,z\,\lesssim\,0.45$ with high-quality COS FUV spectra, which we used to characterize the intrinsic scatter in the gap between low-redshift AGN and the edge of the Ly$\alpha$ forest detected toward them.

We constructed the observed cumulative distribution of the difference between AGN/QSO systemic redshifts and their highest-redshift Ly$\alpha$ absorption line with $\log N_{\rm{\ion{H}{i}}}/{\rm cm^{-2}} > 12.6$, $\Delta z = z_{\rm{sys}} -$ max($z_{\rm{Ly\alpha}}$), and we use the 95\% confidence interval of this distribution as a robust redshift constraint (see equation (1)) when combined with an object's UV-detected max($z_{\rm{Ly\alpha}}$). We explored the systematics of our blazar redshift constraint technique and determined that the systematic uncertainty coming from any differences in the frequency of associated absorption and/or the proximity effect between blazars and typical AGN is sub-dominant compared to the statistical uncertainty on $\Delta z$.

We extended the UV spectral coverage of 1ES 1553+113 with new COS NUV data to confirm its highest-redshift Ly$\alpha$ absorption line at $z= 0.413$, and we applied our Ly$\alpha$-forest-based redshift estimation technique to this blazar. Our redshift constraint for 1ES 1553+113 of $z = 0.408\textrm{--}0.436$ improves confidence in its inferred redshift upper limit and confirms its membership of a galaxy group at $z = 0.433$.

We applied our redshift constraint technique to nine other bright blazars that have archival \textit{HST} UV spectra, most of which are BL Lac objects and key targets for studies of both the hot WHIM and the EBL. Our inferred redshift constraints for these blazars improve the redshift estimates of two (1ES 1118+424 and S5 0716+714) and are consistent with previous redshift estimates for the rest. We have shown that our technique -- based purely on the edge of the observed Ly$\alpha$ forest -- is able to significantly improve redshift constraints for featureless blazars.

Indeed, the most X-ray bright blazars are BL Lac objects \citep[e.g.,][]{Sharma21}, of which a significant portion have unconstrained redshifts as a result of their featureless spectra \citep[][]{Bregman15}, demonstrating the importance of accurate redshift constraints for featureless blazars to interpret future studies of the hot WHIM. Another potential application of our technique that we did not explore in this work is constraining the redshifts of gamma-ray bursts (GRBs) hosted by faint galaxies, which may also be useful probes of the WHIM \citep[][]{Walsh20}.

Our results emphasize the need to obtain further UV and X-ray spectroscopy of blazars and AGN -- complemented with deep galaxy surveys -- in order to improve our understanding of the WHIM and its role in galaxy evolution. \textit{Given the state of current UV facilities, the most immediate need is complementary UV spectra of X-ray bright AGN and blazars in order to fully exploit future possible detections of the hot WHIM in X-ray absorption toward these sightlines.}

\section*{Acknowledgements}

We thank the reviewer for their constructive feedback which helped improve robustness of the results. This research was funded by a National Aeronautics and Space Administration (NASA) grant through HST-GO-15835. Support for program \#15835 was provided by NASA through a grant from STScI, which is operated by the Associations of Universities for Research in Astronomy, Incorporated, under NASA contract NAS 5-26555. Based on observations made with the NASA/European Space Agency (ESA) \textit{HST}, obtained from the data archive at STScI. This work is part of the research programme Athena with project number 184.034.002 and Vici grant 639.043.409, which are financed by the Dutch Research Council (NWO). This research made use of \textsc{Astropy}, a community-developed core Python package for Astronomy \citep{Astropy}, as well as \textsc{NumPy} \citep[][]{Numpy}, \textsc{SciPy} \citep{Scipy}, \textsc{Matplotlib} \citep[][]{Hunter07}, \textsc{emcee} \citep[][]{emcee}, \textsc{lifelines} \citep[][]{Lifelines}, and \textsc{PyQt5}/\textsc{PyQtGraph}. This research has made use of the HSLA database, developed and maintained at STScI, Baltimore, USA. This research has made use of the NASA/IPAC Extragalactic Database (NED), which is operated by the Jet Propulsion Laboratory, California Institute of Technology, under contract with NASA.

\section*{Data Availability}

The NUV data for 1ES 1553+113 presented in this paper is publicly available on the Mikulski Archive for Space Telescopes (MAST, https://archive.stsci.edu/hst/search.php). The FUV data for 1ES 1553+113 and for the AGN/QSOs and blazars used in this paper can be accessed from MAST as well as the \textit{HST} Spectroscopic Legacy Archive (https://archive.stsci.edu/missions-and-data/hsla), which is available in the public domain.
 
\bibliographystyle{mnras}
\bibliography{ms}
\bsp
\label{lastpage}
\end{document}